\documentclass[prb,aps,twocolumn,showpacs]{revtex4}

\usepackage[dvips]{graphicx}
\usepackage{amsmath}

\begin{document}


\newcommand{\etal}{\emph{et al.}}

\renewcommand{\phi}{\varphi}

\newcommand{\iv}{I--V}
\newcommand{\hetemp}{$4.2\unit{K}$}
\newcommand{\be}{\begin{equation}}
\newcommand{\ee}{\end{equation}}
\newcommand{\unit}[1]{\ensuremath{\,\mathrm{#1}}}
\newcommand{\um}{\ensuremath{\unit{\mu m}}}
\newcommand{\uV}{\ensuremath{\unit{\mu V}}}
\newcommand{\mA}{\ensuremath{\unit{mA}}}  
\newcommand{\uA}{\ensuremath{\unit{\mu A}}}  
\newcommand{\ohm}{\ensuremath{\unit{\Omega}}}  
\newcommand{\mV}{\ensuremath{\unit{mV}}}  
\newcommand{\betal}{\ensuremath{\beta_L}}
\newcommand{\betac}{\ensuremath{\beta_c}}

\newcommand{\ich}{$\gamma$ vs.~$f$}
\newcommand{\ic}{\ensuremath{I_c}}
\newcommand{\jc}{\ensuremath{j_c}}
\newcommand{\ib}{\ensuremath{I_B}}
\newcommand{\rb}{\ensuremath{R_B}}

\newcommand{\note}[1]{{\sl #1 }} 

\newcommand{\FQ}{\ensuremath{\Phi_0}}
\newcommand{\RA}{\ensuremath{\Rightarrow}}


\newcommand{\cred}{}
\newcommand{\cblue}{}
\newcommand{\cgreen}{}

\title{Spontaneous creation of discrete breathers in Josephson arrays}
\author{M. Schuster, F. Pignatelli, and A. V. Ustinov}
\affiliation{Physikalisches Institut III, Universit\"at Erlangen-N\"urnberg,
  Erwin-Rommel-Stra\ss e 1, 91058 Erlangen, Germany}

\begin{abstract}
We report on the experimental generation of discrete breather states
(intrinsic localized modes)
in frustrated Josephson arrays. Our experiments indicate
the formation of discrete breathers during the transition from the
static to the dynamic (whirling) system state, induced by a
uniform external current. Moreover, spatially extended
resonant states, driven by a uniform current, are observed to evolve
into localized 
breather states. Experiments were performed on single Josephson
plaquettes as well as open-ended Josephson ladders with 10
and 20 cells. We interpret the breather formation as the result of
the penetration of vortices into the system.
\end{abstract}

\maketitle

Discrete breathers (DBs) are spatially localized, time-periodic
excitations of spatially uniform nonlinear lattices. After initial theoretical
studies\cite{aubry97:_breat,flach98:_discr}, they were observed in
experiments in a variety of
systems\cite{
swanson1999,schwarz1999,trias,binder00:_obser}.
In arrays of superconducting Josephson junctions,
so-called Josephson ladders (JL), DBs can be excited in a
controlled way\cite{trias,binder00:_obser,mazo99:_discr}.
Experiments carried out in JLs ranging from one cell up to several
tens of cells revealed a large variety of possible DB states\cite{binder02:_explor_josep}, studied
their region of existence and the resonant interaction of the
nonlinear DB excitation with small-amplitude linear waves\cite{schuster01:_obser_josep,pignatelli03:_obser_josep}.

DB states in JL are usually excited from the
uniform static state of the lattice by using additional currents (i.e., local forces)
which are removed after the DB creation. This well-controlled procedure
helps to study DB properties systematically.
 In this paper we report that DB can be also excited without
 applying any local forces to the lattice. 
We present several scenarios where DB
states are observed to form spontaneously. The DB formation may occur
 either from the initially static 
 system state, or as
a ``breakdown'' of spatially extended whirling states that are in
 resonance with the cavity eigenfrequency.

\begin{figure}[bt]                                                      
  \includegraphics[width=8cm]{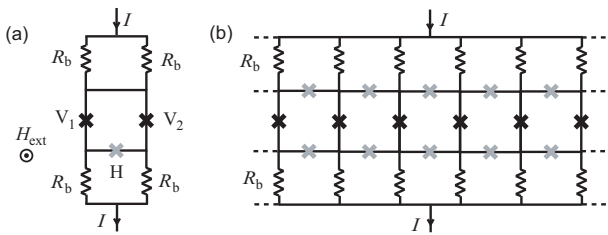} 
  \caption{\label{fig:system} Sketch of the single cell system (a) and
    of the Josephson ladder (b). Josephson junctions are indicated by
    crosses, superconducting leads by straight lines. The homogeneous
    bias current, $I$, is 
    injected through resistors $\rb$.}
\end{figure}

Within the RCSJ (resistively and capacitively shunted junction) model
\cite{barone}, a Josephson junction (JJ) has a simple mechanical
analog, which is a 
pendulum. In this analog, the difference $\phi$ of the
superconducting phases across the JJ corresponds to the tilting
angle of
a pendulum. The current sent across the JJ is
represented by a torque $\gamma$ applied to the pendulum. Both
JJ and pendulum are subject to a velocity dependent friction
force $\alpha\dot\phi$ with damping constant $\alpha$. Finally, the JJ
or pendulum phase moves in a cosine potential due
to the Josephson energy or gravitation, respectively, which leads
to a force proportional to $\sin\phi$.
The dynamical equation governing the phase motion can in either case  be
written in normalized units as
\begin{equation}
  \label{eq:sjjdyn}
  \ddot\phi+\alpha\dot\phi+\sin{\phi}=\gamma.
\end{equation}
The (approximate)
time-averaged solutions $\langle\phi\rangle$ and
$\langle\dot\phi\rangle$ of this equation may have several simple forms.
For $\gamma>1$, Eq.~(\ref{eq:sjjdyn}) has only a \emph{whirling} solution with
$\langle\dot\phi\rangle =
\gamma/\alpha\Leftrightarrow\langle\phi\rangle = (\gamma/\alpha) t$. For
the intermediate range  
$4\alpha/\pi\le\gamma\le1$, a \emph{static} solution with
$\langle\phi\rangle=\arcsin\gamma$ coexists with the whirling
solution. 
For $\gamma<4\alpha/\pi$, only the static solution survives. 
Small-amplitude  \emph{oscillations} of $\phi$ may be considered as
perturbations to the
time-averaged solutions.
Due to the Josephson law, the time derivative of the phase is
proportional to the voltage drop across the JJ. For the whirling
solution, the normalized voltage $v$ is therefore proportional to the current,
$v=\langle\dot\phi\rangle=(1/\alpha)\gamma$, and the inverse of the
damping $\alpha$ is simply the (normalized) ohmic resistance $r$ of
the JJ, $r=1/\alpha$. The real
current $I$ is normalized to the maximum current allowing for the static
state $I_c$, hence $\gamma=I/I_c$, for convenience. 

A JL is a single row of a two-dimensional rectangular Josephson
junction array (see Fig.~\ref{fig:system}b). The JJs are
connected via superconducting 
leads. JJs on the spars of the ladder will be referred to as
the ``vertical'' junctions, those along the rungs as
``horizontal'' junctions. The vertical JJs are all identical and
have an
area of $A_v$, 
likewise all horizontal JJs are identical but have an area
$A_h$. A single cell of the ladder consists
of two 
vertical and two horizontal
JJs. Many of the properties from the extended Josephson ladder are
preserved in a single cell with only one horizontal JJ
(Fig.~\ref{fig:system}a). This system is the simplest
possible Josephson array permitting DB-like excitations.
The anisotropy $\eta$ is defined as the ratio of the
horizontal and vertical JJ areas, $\eta=A_h/A_v$. The JJ
phases inside a cell are coupled due to the magnetic flux quantization
and via a circulating inductive
current. Every cell has a self-inductance $L$. 
A simple model\cite{grimaldi96:_flux} 
describing the phase dynamics of the JL 
neglects mutual inductances between cells and
is similar to the discrete sine-Gordon model.

The vertical JJs form an array of rotors (pendulums). 
These JJs can be driven by a uniform bias current. Technically,
this is achieved by using a single current source which feeds a comb
of parallel resistors $\rb$ (see Fig.\ref{fig:system}b). If the
resistance $\rb$ is large enough, the currents through all vertical branches
are identical.  If the normalized
current $\gamma$ per vertical JJ is large compared to unity, all vertical
JJs rotate. This is the so-called homogeneous whirling state
(HWS). In contrast, if the current is small,
all JJs remain in the static state (SS). In the region
$4\alpha/\pi<\gamma<1$, the ladder may have some vertical JJs in the
whirling state, while the remaining JJs are in the static
state. In the standard terminology, this localized state is called roto-breather
state (RB). Because of to the JL geometry, localized voltage states
always have to conform to Kirchhoff's voltage law. Hence a single
vertical rotating 
JJ has to be accompanied by one or more horizontal JJs in
the whirling state. A large variety of different RB states is
possible, as it has been observed in experiment\cite{peter2000,binder02:_explor_josep}.

The SS of the ladder is influenced by
an externally applied magnetic flux $\Phi_\text{ext}=B_\text{ext}A$
(due to a magnetic field $B_\text{ext}$ 
threading the cell area $A$). The magnetic field leads to a screening
(Meissner) current flowing around the JL. The field may penetrate into the ladder, leading to vortices in the
screening current distribution. 
A measurement of the maximum critical current
$\gamma_c$ of the S state reveals information
on the magnetic field penetration into the ladder.
In Ref.~\onlinecite{binder00:_exper}, the existence region of the S
state of a Josephson ladder was investigated and compared to
theoretical predictions\cite{barahona98:_super}. Deviations were found
for close to half-integer frustration and attributed to the presence
of meta-stable states.
In this paper, we present similar measurements of $\gamma_c$
vs. magnetic field and will argue that they are related to the
creation of DBs. 

When the ladder is in the HWS, all vertical JJ phases rotate at
an identical average frequency $\Omega$, while the horizontal
JJs do not rotate. This is accompanied by a
uniform voltage drop across the vertical JJs. 
Superimposed phase oscillations are strongly enhanced if the
rotation frequency $\Omega$ matches one eigenfrequency of the
cavity  formed by the 
finite ensemble of cells. The enhanced oscillations lead also to an
enhanced damping. 
A suitable experiment is to measure the
differential resistance (which is inversely proportional to the
damping) of the JL. It is defined as the ratio of 
voltage change induced by current variation. The matching
frequencies (voltage positions) are characterized by steps in the
current vs.~ voltage characteristics.
Such resonances of the HWS with
cavity modes have been studied by Caputo \etal\cite{caputo1999}. A
similar behavior was observed when frequency components of a localized RB
state coincide with one of the cavity
eigenfrequencies\cite{schuster01:_obser_josep,pignatelli03:_obser_josep}. We will show
experiments where the HWS resonances are found to enforce the creation
of symmetry-broken RB states. 


We performed the first group of experiments on a single cell, having the central hole area $A=4\times4$
$\mu$m$^{2}$, with three small underdamped Nb/Al-AlO$_{x}$/Nb
JJs \cite{hypres}, Fig.~\ref{fig:system}a. The
vertical JJs have an area $A_{v}=6\times6$ $\mu$m$^{2}$, while the
horizontal JJ has an area $A_{h}=3\times4$ $\mu$m$^{2}$. The
anisotropy of the system is then $\eta=A_{h}/A_{v}=0.33$. At $4.2\unit{K}$ the
parameter of self-inductance evaluated by measurement of a SQUID of the 
same geometry, is $\beta_{L}=2\pi L I_{cv}/\Phi_{0}=0.98$.  Measurements were
done in a temperature range between $4.2\unit{K}$ and $6.3\unit{K}$. At each temperature position the
damping $\alpha=\pi\gamma_{r}/4$ and the critical current density \jc\
are evaluated from the current-voltage characteristic of HWS. At
$T=4.2$ K, $\alpha\simeq 0.03$ and $\jc=126$ A/cm$^{2}$. The bias
current $I$ is introduced via two parallel external resistors,
$\rb=1.5$ k$\Omega$, see Fig.~\ref{fig:system}, so that the bias currents
on the two vertical branches are equal. For convenience, the
experimental data are shown in the normalized units 
of the bias current $\gamma=I/I_{cv}$ and frustration
$f=\Phi_\text{ext}/\Phi_{0}$, where $\Phi_{0}$ is
the flux quantum.

\begin{figure}[tb]
  \includegraphics[width=8cm]{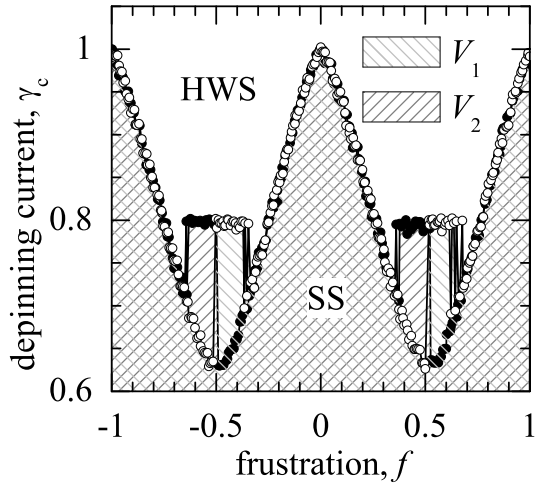}
  \caption{\label{fig:ih}Dependence of the depinning currents
    $\gamma_{c}^{1}$ and $\gamma_{c}^{2}$ 
    on the external frustration, shown
    respectively by full and open circles. The hatched
    patterns show
    the regions where junctions $V_{1}$ and
    $V_{2}$ are in the SS.} 
 \end{figure}

We ramped up the bias current $\gamma$ and registered a transition of
the system to a resistive state by detecting a non-zero voltage at
either junction V$_{1}$ (this critical value of $\gamma$ we call
$\gamma_{c}^{1}$) or junction V$_{2}$ (we call it $\gamma_{c}^{2}$). 
The measured dependence of $\gamma_{c}^{1}$ and $\gamma_{c}^{2}$ on the external magnetic field is, as expected
\cite{barone}, periodic, see Fig.~\ref{fig:ih}. Around $f=\pm0.5$, the behavior of the
two vertical JJs is found to differ and spontaneous formation of
broken-symmetry states \cite{pignatelli03:_obser_josep} is observed. In
Fig.~\ref{fig:ih} the experimental data at $T\simeq 5.5$ K are
shown. In the range of frustration $-0.7\lesssim f\lesssim -0.5$ and
$0.3\lesssim f\lesssim 0.5$, where large clockwise mesh currents
are induced, the junction V$_{1}$ remains in the SS while
the junction V$_{2}$ switches 
to the resistive state. The junction V$_{1}$ switches to the resistive
state at a higher current about $\gamma\simeq 0.8$ that limits the
existence of this broken-symmetry 
breather state \cite{misha2002,pignatelli03:_obser_josep}. As expected, the maximum value of the bias current that allows for the existence of the breather state does not depend on the frustration\cite{pignatelli03:_obser_josep}. In the
range $-0.5\lesssim f\lesssim -0.3$ and $0.5\lesssim
f\lesssim 0.7$, large counterclockwise mesh currents are induced. The
junction V$_{1}$ switches first and the second possible breather state is
spontaneously induced. The range of frustration, where the 
magnetic field induced breather states are observed, broadens with temperature.

\begin{figure}[tb]
  \includegraphics[width=7cm]{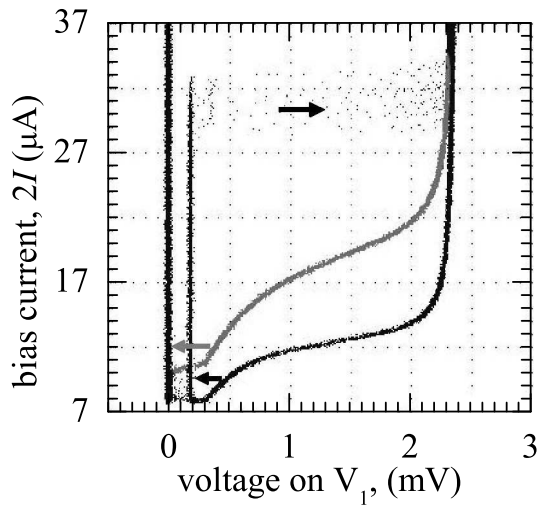}
  \caption{\label{fig:reso}Dynamic switching from the resonance of the HWS (black dots, with a resonant step at about 0.2 mV) to the broken symmetry state (gray points); $f=-0.41$}
\end{figure}

At temperatures higher than $4.2\unit{K}$, the HWS current-voltage
characteristic in presence of external frustration showed the
well-known resonance due to the interaction with the electromagnetic 
oscillatory modes of the cell \cite{barahona1997,caputo1999,caputo2001}. At
$T\simeq5.5\unit{K}$, for 
some values of frustration about 0.5, the interaction of the HWS
with the oscillatory modes of the cell led the system
to switch from the top of the resonance towards a RB state,
see Fig.~\ref{fig:reso}.
 
\begin{figure}[ht]
  \begin{center}
    \includegraphics[width=8cm]{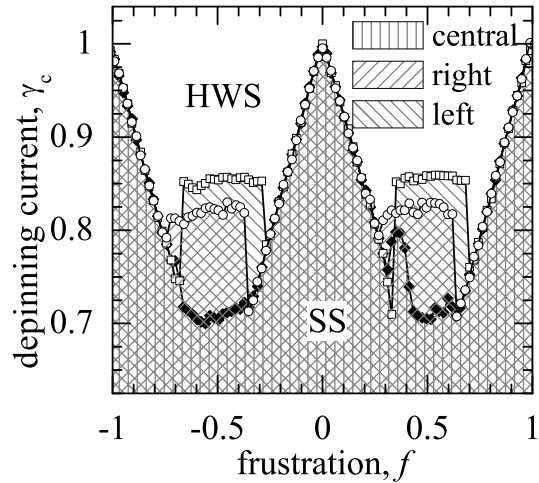}
    \caption{Dependence of the left (open squares, $\gamma_c^1$),
      central (filled 
      diamonds, $\gamma_c^{11}$), and right (open circles,
      $\gamma_c^{21}$) vertical junction depinning currents  
      on the frustration $f$.}
    \label{fig:ichN20_br}
  \end{center}
\end{figure}

The second group of experiments has been performed with
open-ended JLs.
Measurements of the critical current $\gamma_c^i$ needed to force the
vertical JJ $i$ to the finite voltage state are presented in
Fig.~\ref{fig:ichN20_br}. The sample consisted of 20 cells,
with parameters $\eta=0.5$, $\betal\approx1.2$ and $\alpha\approx0.03$. The
sample layout is similar to the one presented in Ref.~\onlinecite{binder00:_exper}. The
measurement was performed at a temperature of
\hetemp.

The pattern of the depinning current $\gamma_c^i$ is periodic in
frustration, with the period corresponding to a magnetic
flux of $\Phi_0$ per cell. 
Around integer values of $f$, the critical currents for left, center, and right positions in the
ladder coincide. After the depinning, for $\gamma\ge\gamma^i_c$,  the
HWS is formed. For $f\sim0.5$, the critical currents $\gamma_c^i$
differ 
for voltages measured on different positions in the ladder. At
$-0.35<f<-0.25$, the right and central 
vertical JJs of the ladder start to rotate at a lower value of
$\gamma$ than the left part
of the ladder. For $-0.65<f<-0.35$, the central part depins at the
lowest value of the current, the right part at an intermediate
current, and the left part at the highest value of the current. For
$-0.75<f<-0.65$ the center and left parts depin for a low current and
the right part depins at a higher current.

By definition, a breather state is formed when one part of the
vertical JJs in 
the ladder rotates while the rest of it remains in the SS. 
Hence in
the described frustration regions around $f\sim0.5$, breathers are
found. They form solely
due to an external magnetic field. 
Because of the limited number of voltage probes in our long ladder, we
were not able to exactly
identify the precise number of rotating JJs in the breather states which were formed, but could clearly
distinguish them from the HWS.

Our observations can be qualitatively explained in the following
way. When the HWS has formed for large $\gamma$ and finite
frustration $f$, magnetic flux penetrates into the system via the
boundary cells. When $\gamma$ is reduced to zero, the ladder settles
in the SS, which may be a meta-stable state or the system
ground state. Above some critical value of the frustration $f$, this
state consists of a nonzero amount of
vortices\cite{barahona98:_super}. For small bias
currents $\gamma$, the vortices are pinned due to the
lattice discreteness. However, when the vortices start to move, they
may ultimately bump into the ladder boundary. This will eventually force
the boundary JJ to the whirling state, with the rest of the
ladder remaining in the static state. The low value of $\eta$ is
crucial for this behavior. Preliminary numerical
simulations have shown this kind of behavior for typical experimental
parameters\cite{schuster:_gener_RB}.
For longer systems, it is also expectable
and observed that moving vortices may be expelled via one of the horizontal
JJs before reaching the boundary. When a vortex passes
``through'' a horizontal junction, this phase may be left rotating,
again leading to the formation of a RB. A series of DBs may be left
behind a moving vortex. This has been observed also in the simulations
presented in Ref.~\onlinecite{trias02:_inter}. We conclude that the depinning
transitions for JLs with low anisotropy $\eta$ can lead to the
formation of \emph{localized whirling (RB) states}.


\begin{figure}[tb]
  \begin{center}
    \includegraphics[width=8cm]{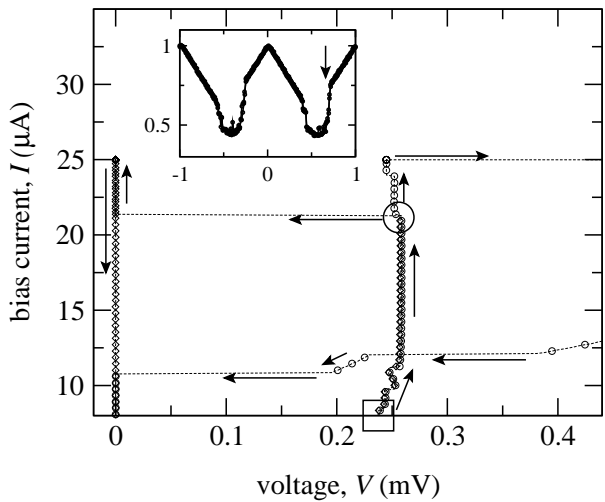}
    \caption{Current-voltage curve of a 10-cell JL indicating the creation of a
      breather state from a whirling mode resonance for a frustration
      $f=0.66$.
      The current was swept from $I=8\unit{\mu A}$ to $I=25\unit{\mu
      A}$ and back.
      Both the central (circles) and the right boundary
      (diamonds) voltages are
      plotted. Arrows indicate directions of the voltage
      switching and the path which is taken during the current
      sweep. The square indicates the initially prepared resonant HWS
      ($I=8.4\unit{\mu A}$). The circle indicates
      the position at which the symmetry-broken state is formed ($I=21.4\unit{\mu A}$). The
      inset shows the dependence of the center junction depinning
      current on frustration, the arrow marks the frustration at which the
      current-voltage curve was taken.}
    \label{fig:iv_br}
  \end{center}
\end{figure}

We also observed the creation of a RB state from a
\emph{resonant} HWS in JLs.
These experiments were performed with a 10 cell open-ended JL of 
anisotropy
$\eta=0.49$. Other parameters during the measurement were
$\betal\sim0.35$ and $\alpha\sim0.03$. 
In the presence of the external magnetic field, we observe
resonant steps on the current-voltage characteristics of the
ladder, as shown in Fig.~\ref{fig:iv_br}. They are occur when the
vertical JJ 
rotation frequency $\Omega$ matches one of the  cavity eigen-frequencies, as
discussed in Ref.~\onlinecite{caputo1999}.
 When the homogeneous current is increased at a resonance, the
amplitude of the cavity oscillations increases. If the driving
current becomes
too large, the resonant state cannot be maintained anymore and the
non-resonant HWS (now with a larger rotation frequency)
is retained. This hysteretic process is usually repeatable for an
arbitrary number of times.  However, in some rare but reproducible
cases, we observed 
a transition from such a
resonant, spatially homogeneous state towards a localized RB state. The
localized state RB can be identified as before by a zero voltage drop across
some vertical 
JJs in parallel with a finite voltage drop across another group
of vertical
JJs. 
In Fig.~\ref{fig:iv_br}, the voltage switching, and hence
the RB formation happens in the region emphasized by the circle,
around $I=21.4\unit{\mu A}$. During further increase of the current
the RB voltage stays constant, apart from some tiny jumps. We thus observe
here that a
\emph{resonant} RB state is formed from the resonant HWS. 
Ultimately,
the RB voltage switches to a high value (at
$I\approx25\unit{\mu A}$). Here the breather 
state is still present, but is non-resonant. 
The tiny voltage jumps along the RB
resonance may indicate either transitions between different excited
modes or 
a change in 
the RB size, which could not be detected in this experiment due to the
limited number of 
voltage probes.
The described transition occurs for a very narrow range of magnetic
field. 

In summary, we reported here on the spontaneous creation of discrete
breather states in single Josephson plaquettes and Josephson
ladders. The excitation of breather states is observed during the
transition from the static to the whirling state of the array,
resulting from a ramp of an externally applied homogeneous bias
current. In another regime, the breather states are created from the
resonant uniformly whirling state of the array. The spontaneous
creation appears in the presence of an
external magnetic field close to half-integer frustration, which leads
to large screening currents. The observed behavior may be related to
the appearance of symmetry-broken states in two-dimensional Josephson
arrays\cite{abraimov99:_broken_josep}. An analysis of the
described phenomena could provide more insight in the physical
relevance of localized voltage states in
Josephson coupled systems.





\end{document}